\documentclass{aa}  
\usepackage{amsmath}
\usepackage{graphicx}
\usepackage{txfonts}
\usepackage{hyperref}
\hypersetup{
     colorlinks=true,
     linkcolor=blue,
     filecolor=blue,
     citecolor = blue,      
     urlcolor=blue,
     }

\newcommand* {\Msun} {\mbox{M$_\odot$}}
\begin{document}

   \title{Analogues of the Milky Way-Sagittarius interaction in the TNG50: effect on the Milky Way}

   \titlerunning{MW-Sgr -like interactions in TNG50}
   \authorrunning{Marcin Semczuk et al.}

   \author{Marcin Semczuk
          \inst{1,2,3}
          \and
            Teresa Antoja
            \inst{1,2,3}
          \and
            Alexandra Girón-Soto
            \inst{1,2,3}
        \and Chervin F. P. Laporte
        \inst{4,2,5}
            }

   \institute{Departament de Física Qu\`antica i Astrof\'isica (FQA), Universitat de Barcelona (UB),  c. Mart\'i i Franqu\`es, 1, 08028 Barcelona, Spain
         \and
Institut de Ci\`encies del Cosmos (ICCUB), Universitat de Barcelona (UB), c. Mart\'i i Franqu\`es, 1, 08028 Barcelona, Spain
        \and
Institut d'Estudis Espacials de Catalunya (IEEC), Edifici RDIT, Campus UPC, 08860 Castelldefels (Barcelona), Spain
        \and
LIRA, Observatoire de Paris, Universit\'e PSL, Sorbonne Universit\'e, Universit\'e Paris Cit\'e, CY Cergy Paris Universit\'e, CNRS, 92190 Meudon, France
        \and
Kavli IPMU (WPI), UTIAS, The University of Tokyo, Kashiwa, Chiba 277-8583, Japan
             }

   \date{submitted to A\&A}

 
  \abstract
{The Sagittarius Dwarf Galaxy is undoubtedly being disrupted in the tidal field of the Milky Way. The Milky Way disc is also found to be in a state of disequilibrium. The role of the Sagittarius in driving or contributing to this disequilibrium has been extensively investigated. Most of these studies, however, assume an initially near-equilibrium disc. It was also hypothesized that the passage of Sagittarius could have increased the star-forming activity in the Solar Neighbourhood.}
{We check whether galaxies that have undergone cosmological evolution are affected by interactions analogous to those between the Sagittarius and the Milky Way.}
{We use the high-resolution simulation TNG50 to look for pairs similar to the Milky Way and Sagittarius. We search within redshift $z=1$–$0$ for discs from the MW/M31 sample that interacted with a satellite more massive than $10^{10}$ \Msun, had a pericenter smaller than 50 kpc, and was on an approximately polar orbit. We exclude cases where, within 1 Gyr of the pericenter, a similar interaction occurred.}
{In $\sim90\%$ of cases, a passage of the Sagittarius analogue had no significant effect on either the vertical velocity field of the disc or the star formation history. A response in vertical stellar kinematics can be found mostly in cold discs (with mean $\sigma_z<80-90$ km/s) and mildly correlates with the strength of interactions.
For star formation, the studied interactions had an effect only when little to no star formation was ongoing prior to the interaction, often due to previously disturbed star-forming discs, e.g., from AGN activity.}
{Our results indicate that stellar discs in TNG50 are frequently vertically perturbed preceding pericenter passages of Sagittarius analogues. Future studies using other simulations and extragalactic surveys will help establish whether vertical disequilibrium is a common feature of disc galaxies or an artifact of the specific setup studied.}


   \maketitle
%

\section{Introduction}

\begin{figure*}
\centering
\includegraphics[width=18.4cm]{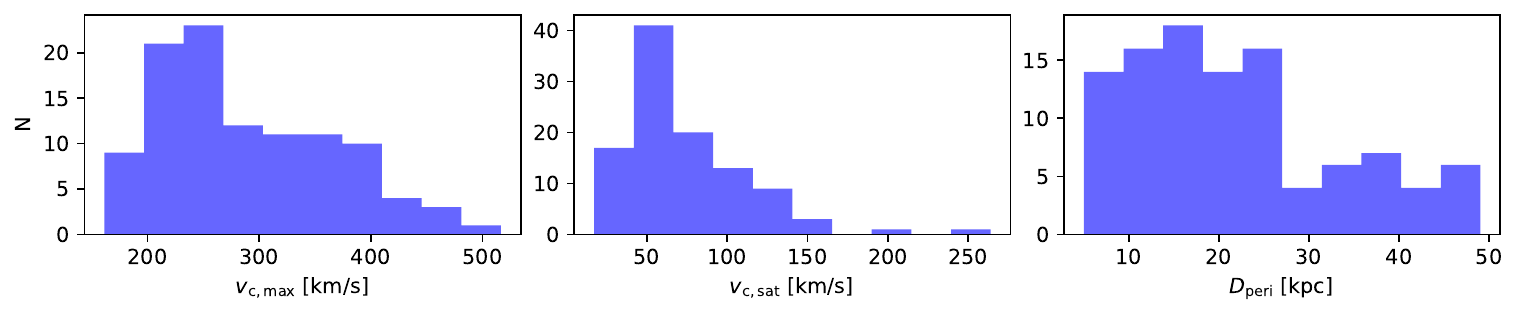}
\caption{Histograms of basic properties of hosts and satellites from our fiducial sample of MW-Sgr -like interaction analogues. Left: Distribution of maximum circular velocities of hosts. Middle: Distribution of maximum circular velocities of satellites. Right: Distribution of pericentric distances.}
\label{sample}%
\end{figure*}

The Sagittarius Dwarf Spheroidal Galaxy (Sgr) is a close neighbour of the Milky Way (MW) that is currently being tidally disrupted by it \citep{Ibata1994,Ibata1995}. The disruption is evidenced by a spectacular stellar stream spanning across the sky \citep{Newberg2002,Majewski2003,Belokurov2006}. Thanks to its proximity and the resulting wealth of data, the MW–Sgr interaction has served as a testbed for studying how satellite galaxies interact with larger late-type hosts for nearly three decades.

Early models of the interaction focused on inferring properties of the MW's gravitational potential (e.g. \citealt{Law2010}) or the progenitor of Sgr (e.g. \citealt{GF1999,Lokas2010}). While these topics are still being explored (e.g. \citealt{Vasiliev2021,DelPino2021,Lokas2024}), new avenues of research have also emerged, such as studying the properties of the stream itself (e.g. \citealt{Ramos2022}). The most recently popular focus, however, was sparked by the discovery of the phase-space spiral in Gaia DR2 by \cite{Antoja2018}, which led to a wave of interest in the effect of Sgr on the MW disc.

\cite{Purcell2011} showed in $N$-body simulations that Sgr could have induced ring-like overdensities in the MW stellar disc. \cite{Gomez2013}, also using $N$-body models, showed that Sgr is capable of inducing vertical waves in MW discs, visible as corrugations in both the density and vertical velocity field. Also using collisionless models, \cite{Laporte2018} studied how Sgr and the LMC can act together to impact the vertical kinematics of the MW disc. Using the same models, they later showed how vertical waves project onto $z$–$v_z$ \citep{Laporte2019} in the form of the newly discovered phase spiral. The interaction with Sgr was also shown to be a possible source of the phase spiral through test-particle simulations \citep{Binney2018}. A summary of recent work on the phase spiral has been compiled in Section 5.3 of \cite{Hunt2025}. Besides the effect on vertical structure and kinematics, \cite{Antoja2022} and \cite{Bernet2025} studied the impact of Sgr in creating spiral arms and planar kinematic waves.

Beyond the possible effects of Sgr on the kinematics and structure of the MW stellar disc, \cite{RuizLara2020} proposed that recurring passages of Sgr could have triggered periods of increased star formation rate (SFR). This claim was based on the coincidence of peaks in their derived star formation history (SFH) with pericenter passages from several Sgr orbit models. \cite{Bhargav2024} tested the feasibility of this scenario by running a suite of hydrodynamical simulations with varying satellite orbits and initial gas, stellar, and dark matter content. Their analysis concluded that the connection between the Sgr's orbit and star formation bursts is likely to be casual.

Most of the work mentioned above that studied the effect of Sgr on the MW disc and SFH has been done within idealized, near-equilibrium frameworks. The emergence of high-resolution cosmological simulations and zoom-ins offers new insights into galaxy interactions similar to those between the MW and Sgr. \cite{Gomez2016} found in the Auriga \citep{Grand2017} simulations that an interaction with a small satellite, amplified by the dark matter halo wake, can induce a corrugation pattern in the disc, claimed to be observed in the MW by \cite{Xu2015}. In \cite{Gomez2017}, using the same suite of zoom-in simulations, they showed that warps in stellar discs, while very common, are likely tidal in origin when seen in all stellar populations. \cite{GarciaConde2022} found phase spirals associated with bending waves in the GARROTXA simulations \citep{Santi2016}, though several agents were identified as possible triggers, making it difficult to establish a single cause \citep{GC2024}. \cite{Grand2023}, using higher-resolution zoom-in SUPERSTARS simulations, showed that phase spirals can last several Gyr after being induced by dark matter halo wakes associated with satellite passages.

In this work, we expand on results obtained in the cosmological framework by reversing the questions posed by previous authors. Instead of searching for symptoms of satellite-disc interactions (e.g. warps, phase spirals), we search the TNG50-1 run of the IllustrisTNG suite \citep{Pillepich2019,Nelson2019,Nelson2019b} for galaxy pairs whose interactions resemble that of the MW and Sgr. Having a reasonably sized sample of such interactions allows us to statistically assess how likely it is for cosmologically evolved discs to be significantly perturbed by similar interactions.

The paper is structured as follows. In Section 2, we describe the selection of our sample from the TNG50 box. Section 3 discusses the effects of satellites on the vertical kinematics of discs. In Section 4, we report on the impact on the star formation history. Section 5 discusses the results in a broader context and summarizes our findings.

\section{Sample selection}
\label{sec:sample}
To search for analogues of MW–Sgr interactions in TNG50, we start from the catalogue of MW/M31 galaxies in this run compiled by \citet{Pillepich2024}. We use the 192 simulated galaxies at $z=0$ that were analyzed by \citet{Engler2021,Engler2023} in the context of satellite systems. To find Sgr-like perturbations acting on these disc galaxies, we define them according to the following four criteria:

\begin{enumerate}
\item The perturber had a minimum total mass of $10^{10}\;\Msun$ while being within a radius of 100 kpc from the host.
\item The perturber had a mass smaller than the host.
\item The perturber had a pericenter passage closer than 50 kpc and no other satellite above the mass of $10^{10}\;\Msun$ had a pericenter within 1 Gyr before or after the snapshot closest to the pericenter.
\item The orbit of the perturber was close to polar within 1 Gyr around the pericenter passage. We define this as passages where the angle between the host's disc angular momentum (measured as total stellar angular momentum within $2 r_\mathrm{h}*$, where $r_\mathrm{h}*$ is the stellar half-mass radius) and the satellite's orbital angular momentum was in the range $60^\circ$–$120^\circ$ during that time window.
\end{enumerate}

We searched the environments of MW/M31 analogues within the redshift range $z=1$ to $z=0$ (corresponding to a time range of 8 Gyr) for interactions matching the above criteria. We found 62 host galaxies that experienced 119 interactions fulfilling these conditions. During the analysis, we noticed that for several host galaxies stellar disc planes were not well aligned with the total stellar angular momentum within $2 r_\mathrm{h}*$.
Visual inspection revealed several reasons affecting the definition of the disc plane in these cases: 
i) very small, compact galaxies with pericenter passages occurring close to $z=1$,
ii) polar ring galaxies (similar to those studied in \citealt{Smirnov2024}) or other rotating structures appearing within the radius of $2 r_\mathrm{h}*$, and
iii) very violent ongoing mergers. As none of these three situations are representative of the MW, we excluded them from our sample, obtaining a fiducial set of 52 host galaxies that experienced 105 pericenter passages similar to those of the Sgr dSph.
The distribution of basic parameters of hosts and satellites from the fiducial sample are shown in Fig.~\ref{sample}.

\begin{figure}
\centering
    \includegraphics[width=.99\columnwidth]{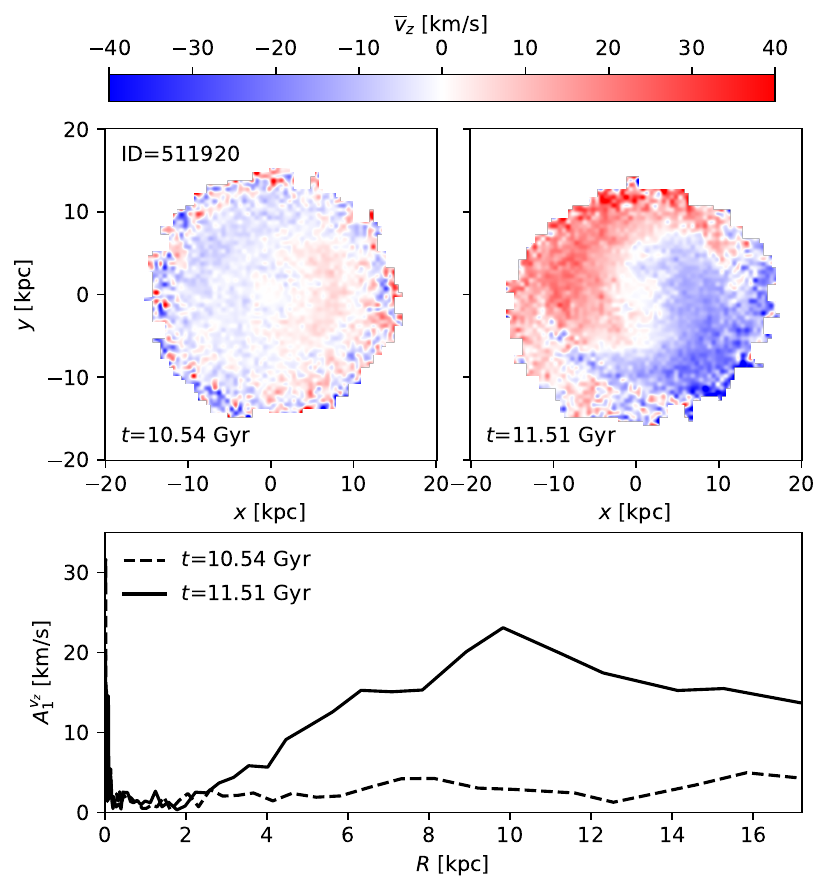}
    \caption{Top: examples of $\overline{v}_z$ velocity fields of a simulated galaxy at times where $m=1$ Fourier amplitude is relatively low (left) and high (right). Bottom: radial profiles of the $m=1$ Fourier amplitude of $\overline{v}_z$: $A_1^{v_z}$ measured for both times. The difference between the values at $R>4$ kpc demonstrates that the amplitude is a good measure of $m=1$ signal in vertical kinematics.}
    \label{fourier}%
\end{figure}

\section{Vertical kinematics}
\label{sect:kin}
\subsection{Fourier analysis}

\begin{figure}
\centering
    \includegraphics[width=.99\columnwidth]{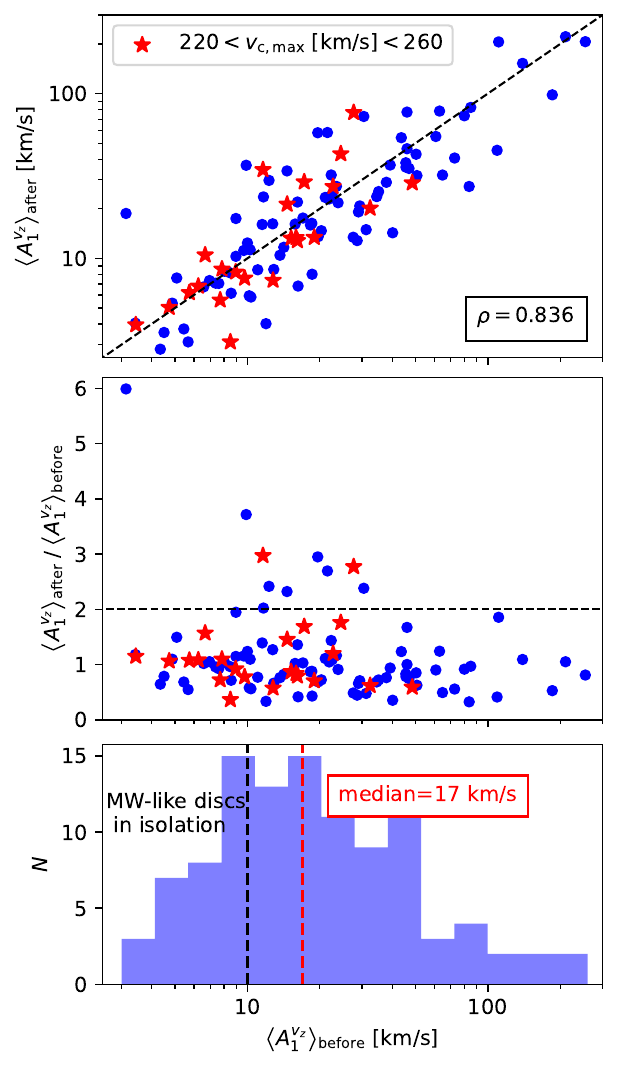}
    \caption{Top: Correlation between the mean amplitude of the $m=1$ mode of $v_z$ before pericenter passage and the mean amplitude of the same mode after pericenter passage. Red stars mark host galaxies with maximum circular velocities closest to the MW. The Spearman correlation coefficient $\rho$ is shown. Middle: Correlation between the mean amplitude of the $m=1$ mode of $v_z$ before pericenter passage and the ratio of the mean amplitude after to before. Bottom: Histogram of the mean amplitude of the $m=1$ mode of $v_z$ before pericenter passage. The median of the distribution is indicated, along with a reference value from \cite{Chequers2017}, measured in MW-like $N$-body discs evolved in isolation.}
    \label{vz0}%
\end{figure}

\begin{figure*}
\centering
\includegraphics[width=18.4cm]{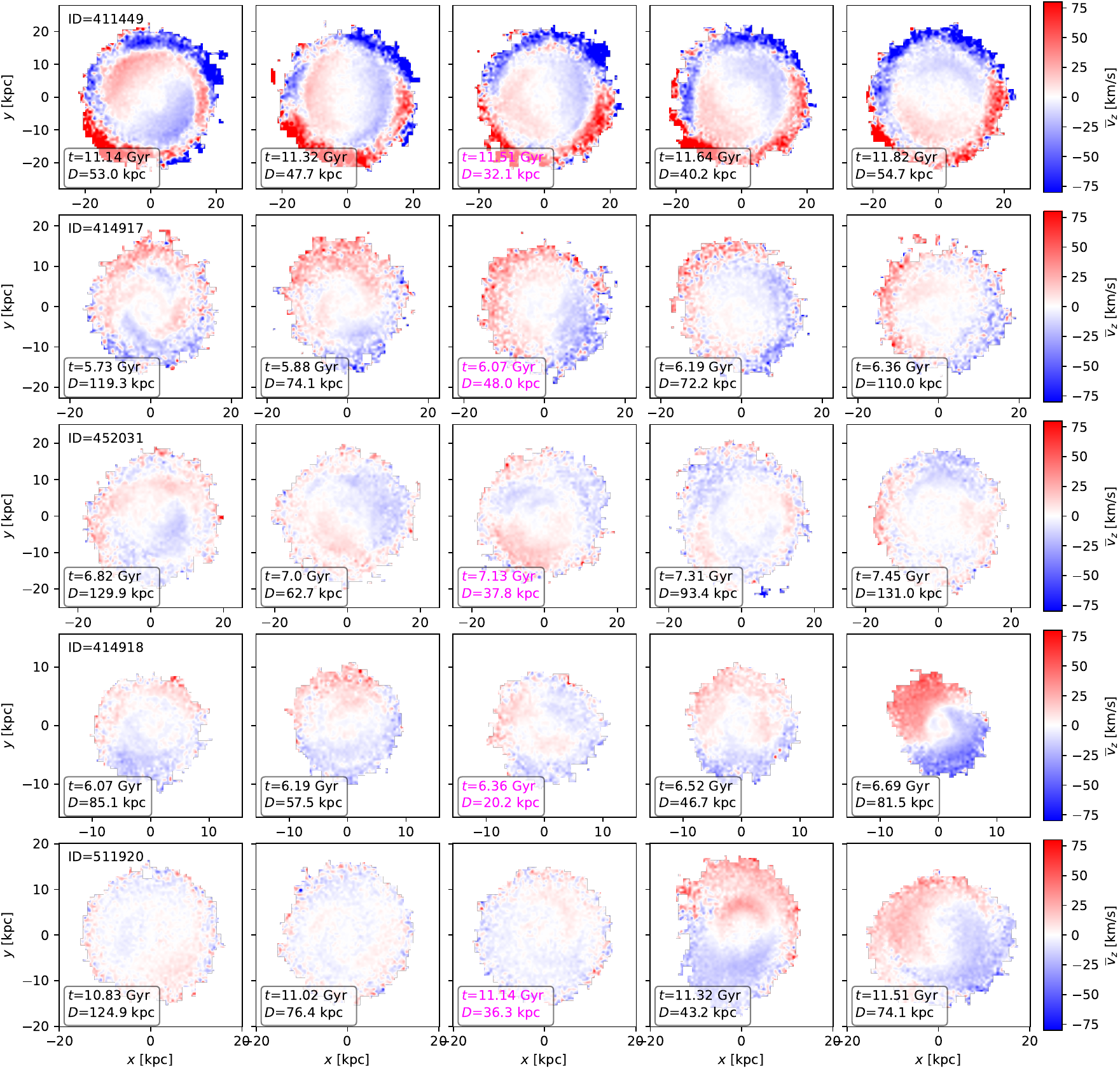}
\caption{Five examples of the temporal evolution of $\overline{v}_z$ fields of MW-like simulated galaxies around the passage of Sgr-like satellites. The middle panels of each row show the snapshot closest to the pericenter. The top three rows show cases where the vertical velocity field is unaffected by Sgr-like interactions, which is the most common behaviour in the studied fiducial sample. The bottom two rows show outlier examples where Sgr-like passage coincides with a growth of the strong $m=1$ signal.}
\label{examples}%
\end{figure*}

\begin{figure*}
\centering
\includegraphics[width=18.4cm]{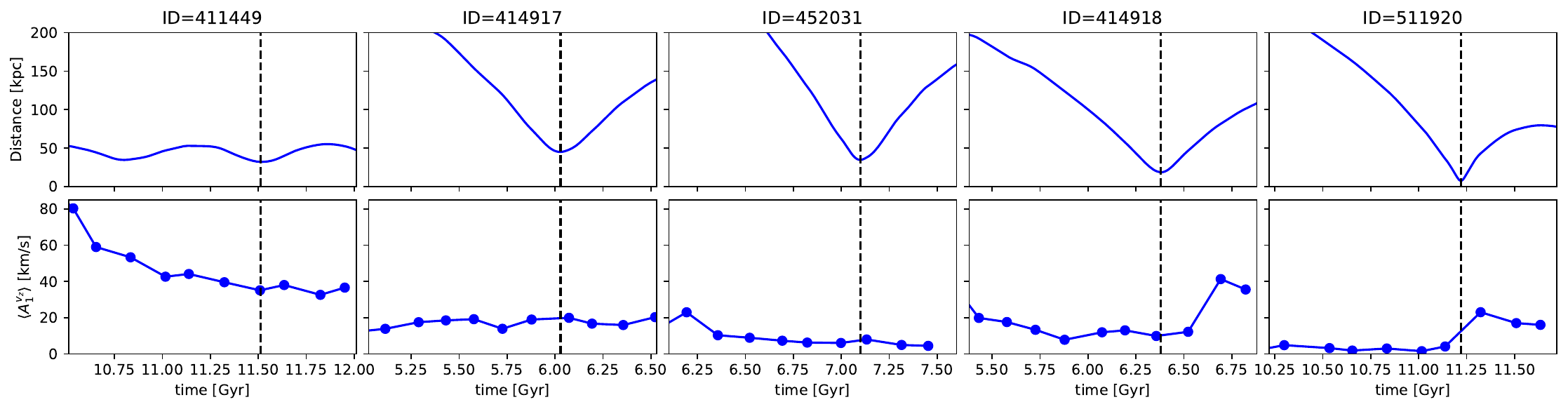}
\caption{Top: Time evolution of the relative distances between the host galaxies and the satellites for the five examples presented in Fig.~\ref{examples}. Bottom: Time evolution of the amplitudes of $m=1$ mode of $v_z$ averaged in the ring of 2 to 6 disc scaleneghs for the same five examples. Dashed lines mark the times of the considered pericenters in both rows.}
\label{orbits}%
\end{figure*}

\begin{figure}
\centering
    \includegraphics[width=.99\columnwidth]{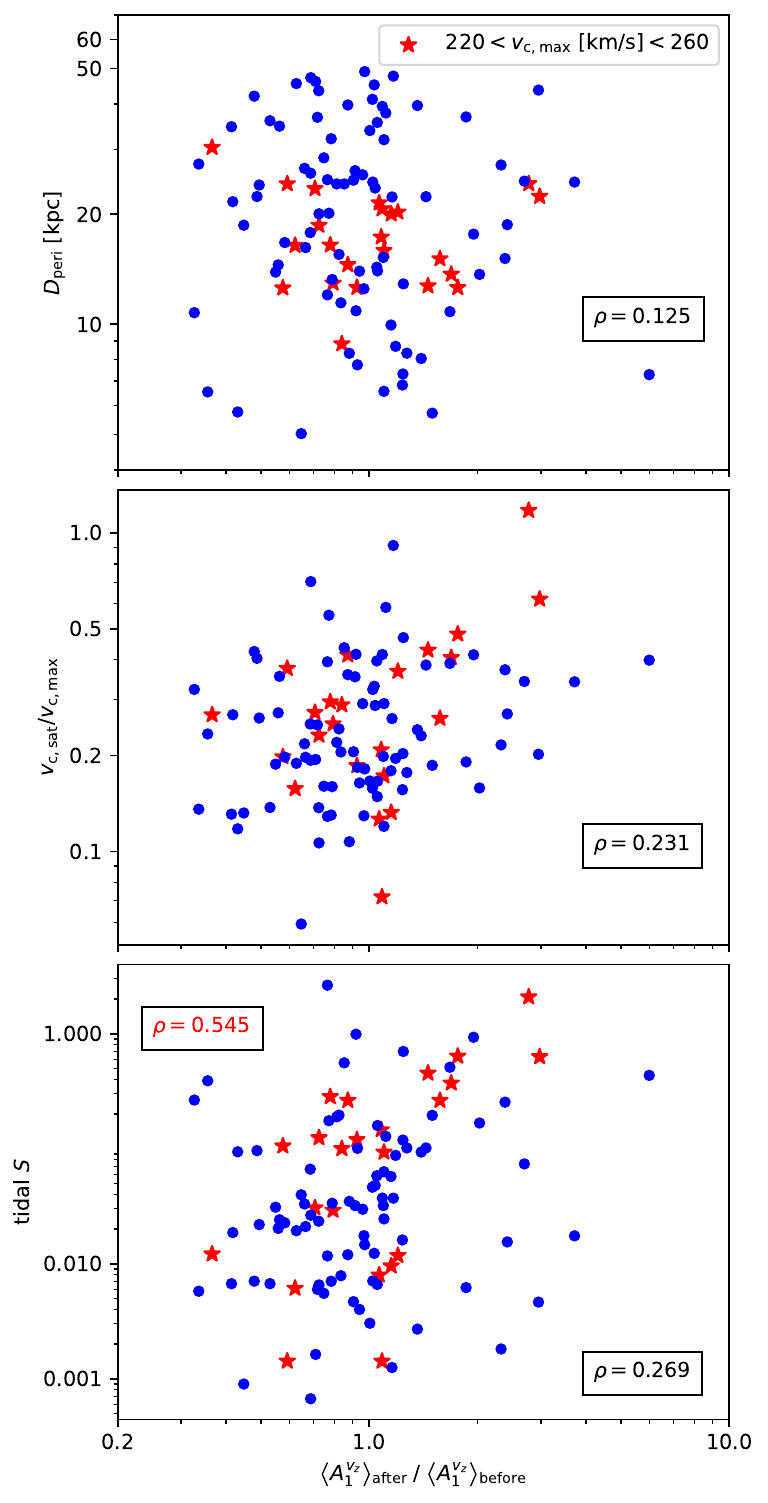}
    \caption{Correlations between the ratio $\left\langle A_1^{v_z} \right\rangle_\mathrm{before}/\left\langle A_1^{v_z}  \right\rangle_\mathrm{after}$ which describes the amplification of the $m=1$ Fourier amplitude around the pericenter of the perturber and pericentric distances $D_\mathrm{peri}$, ratios of the maximum circular velocities of the satellite to the host $v_\mathrm{c,sat}/v_\mathrm{c,max}$ and tidal parameters $S$ as defined in equation~\ref{P}. Red stars mark host galaxies with maximum circular velocities closest to the MW. The Spearman's correlation coefficient $\rho$ is indicated.}
    \label{vz1}%
\end{figure}

\begin{figure}
\centering
    \includegraphics[width=.99\columnwidth]{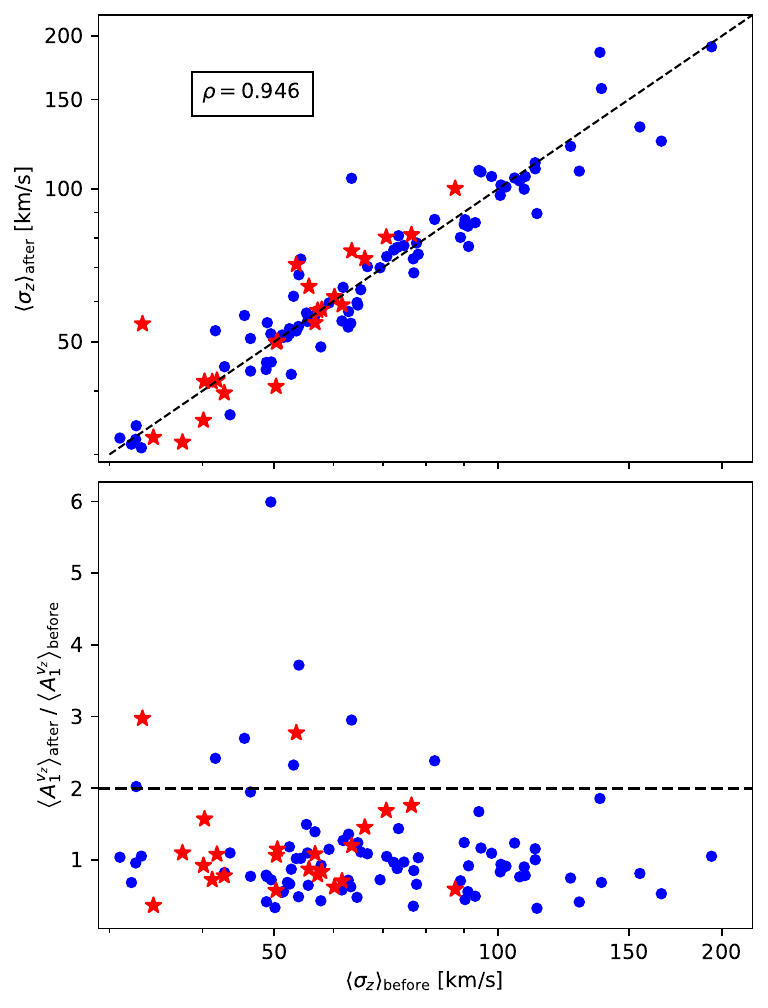}
    \caption{Top: Mean vertical velocity dispersion averaged in 6 snapshots before the pericenter passages vs mean vertical velocity dispersion averaged in 3 snapshots after the pericenter passages. Bottom: Mean vertical velocity dispersion averaged in 6 snapshots before vs the averaged after-to-before ratio of $m=1$ Fourier amplitudes of $\overline{v}_z$. Red stars mark host galaxies with maximum circular velocities closest to the MW. The Spearman's correlation coefficient $\rho$ is indicated.}
    \label{sigma_z}%
\end{figure}

\begin{figure}
\centering
    \includegraphics[width=.99\columnwidth]{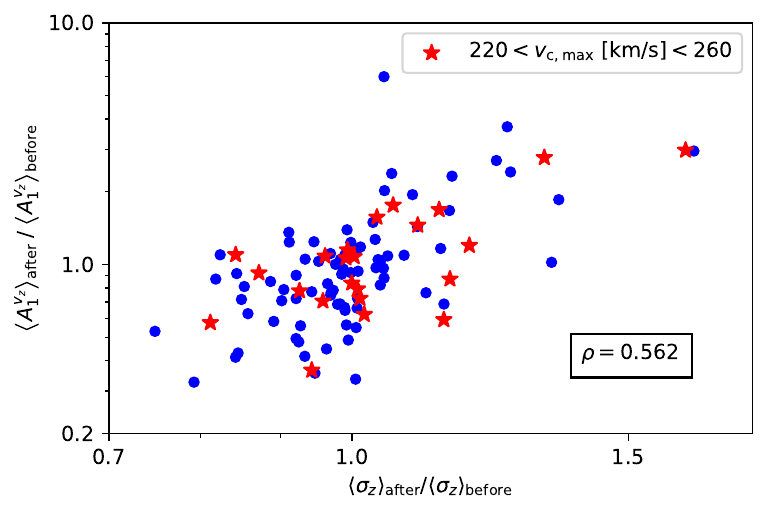}
    \caption{After-to-before ratio of the averaged vertical velocity dispersion against the averaged after-to-before ratio of $m=1$ Fourier amplitudes of $\overline{v}_z$. Red stars mark host galaxies with maximum circular velocities closest to the MW. The Spearman's correlation coefficient $\rho$ is indicated.}
    \label{sigma_v_fourier}%
\end{figure}
To quantify the effect that satellites have on the kinematics of the host discs in our fiducial sample of MW-Sgr analogues, we used a Fourier decomposition of the mean velocity field in the the vertical direction $\overline{v}_z$. Fourier decomposition has been widely used in studying disc features \citep{Sellwood1986} which also includes vertical response of discs to interactions \citep{Chequers2017,Chequers2018,Poggio2021,GC2024}. Here we calculate the $m=1$ amplitude of $v_z$, which describes the S-shaped, 'integral'-sign warps. We calculate it in radial bins according to the following equations:
\begin{align}
a_m=2\sum_{i}^{n}w_i v_{z,i} \cos({m \varphi_i}) \nonumber\\
b_m=2\sum_{i}^{n}w_i v_{z,i} \sin({m \varphi_i}) \\
A_m ^{v_z}=\sqrt{a_m^2+b_m^2}/\sum_{i}^{n}w_i, \nonumber
\end{align}
where $v_{z,i}$ are the velocities in the vertical direction of stellar particles in a given radial bin and $w_i$ are their smooth window function weights. Radial binning and weighting are performed in the same manner as in the Fourier analysis implemented in \texttt{patternSpeed.py} \citep{Dehnen2023}, a code designed to measure the pattern speed of bars from individual simulation snapshots using Fourier decomposition. The use of window-function-based weights and adaptive bin sizes, scaled with the number of particles, yields a smoother radial dependence of the Fourier amplitude for a given quantity, compared to using fixed-size top-hat bins.

To tell whether the fly-by of the Sgr-like satellite made a difference in the vertical velocity field, profiles of $A_1 ^{v_z} (R)$ were calculated in the 6 snapshots preceding the snapshot closest to the pericenter passage in each case, and the 3 following ones. We use more snapshots before to capture the average state of the disc before the interaction. We use less snapshots after the pericenter because we are interested in capturing more instantenous changes, which would otherwise be less visible, if averaged over a greater range.
These intervals correspond to times approximately 600 Myr before and 300 Myr after, with a snapshot cadence of $\sim100$ Myr in the studied redshift range. This cadence also means that the exact time of pericenter can differ from the nearest snapshot by up to 100 Myr. The top and middle panels of Fig.~\ref{fourier} show the $\overline{v}_z$ fields of one simulated galaxy for two example times that had weak and strong $m=1$ symmetry, respectively, which is characteristic of tidal forcing on galactic discs. The bottom panel shows the radial profiles of $A_1 ^{v_z}$ for the these examples which illustrates the robustness of this approach. 

To ensure that the sample of stellar particles used for the Fourier analysis is not contaminated by halo stars or interlopers from the perturber, we made a geometrical cut in $|z|<2 z_\mathrm{s}$, where we define the vertical disc scale $z_\mathrm{s}$ to be  $z_\mathrm{s}=\sqrt{\widetilde{z}^2}$, where $\widetilde{z}$ is the median of $z$ of stars inside $10r_{\mathrm{h}}*$ at the snapshot closest to the pericenter. Other cleaning procedures such as using a different geometrical cut ($|z|<4 z_\mathrm{s}$) or a cut in angular momentum (selecting only stars with $L_z/L>0.7$) led to the same conclusions.

To compress the 1D $A_1 ^{v_z} (R)$ profiles describing the $m=1$ symmetry of the velocity field into individual numbers that can be analysed through correlations, we first averaged their values in the radial range corresponding to the equivalent accessible parts of the MW disc. We defined this range to be between 2 and 6 disc scalelengths $R_\mathrm{D}$, which we estimated by fitting an exponential disc to the surface density profiles in the range\footnote{This range was chosen to exclude the inner bar and bulge regions from the fitting procedures, as their inclusion would bias the measured disc scale lengths. Visual inspection of the sample showed that, beyond $r_\mathrm{h}^*$, the steeper inner slopes, caused by the presence of the bar and bulge, typically taper off, allowing the exponential disc to be accurately fitted beyond this radius.} of 1 to 6 $r_\mathrm{h}*$. The radially averaged amplitudes were later averaged over 6 snapshots before the pericenter $\left\langle A_1^{v_z} \right\rangle_\mathrm{before}$ and 3 snapshots after $\left\langle A_1^{v_z}  \right\rangle_\mathrm{after}$ to describe the average state of the $m=1$ symmetry of $v_z$ field of discs before and after the pericenter passage of the Sgr analogues.

\subsection{Results}
In Fig.~\ref{vz0} we plot the average $A_1^{v_z}$ amplitudes before the pericenters against the average of them after pericenter, ratios after-to-before of $m=1$ Fourier amplitudes and the histogram of average amplitudes before. Since the range of rotation curves in the MW/M31 analogues sample from \cite{Pillepich2024} is quite broad at $z=0$ (values between 150 to 350 km/s), in Fig.~\ref{vz0} and following figures, we highlight with red stars those discs whose rotation curve is closest to the MW (e.g. \citealt{Mroz2019}) by limiting the $v_\mathrm{c,max}$ in the range of 220-260 km/s at the time of the interaction.

The strong correlation between $\left\langle A_1^{v_z} \right\rangle_\mathrm{before}$ and $\left\langle A_1^{v_z}  \right\rangle_\mathrm{after}$ and its proximity to the 1:1 line
indicates that, on average in our fiducial sample, pericenter passages of the Sgr analogues have little effect on the $v_z$ velocity fields. However, some scatter exists on top of this scaling, with 10 out of 105 pericenters having the ratio of $\left\langle A_1^{v_z} \right\rangle_\mathrm{after}/\left\langle A_1^{v_z}  \right\rangle_\mathrm{before}>2$, which is more clear in the middle panel of Fig.~\ref{vz0}. Interestingly, the range of mean $m=1$ amplitudes is quite broad, with the histogram in the bottom panel of Fig.~\ref{vz0} showing its distribution. \cite{Chequers2017} studied spontaneous bending waves in isolated MW-like discs and found the $m=1$ amplitude to be within the limit of 9-10 km/s. The median of the distribution of our fiducial sample equal to 17 km/s exceeds the value from isolated experiments, meaning that MW-like discs in TNG50 are often vertically perturbed, before they experience interactions with Sgr-like satellites.

In Fig.~\ref{examples} we show five examples of behaviour of the vertical velocity fields. The top three rows show the evolution of three galaxies that represent the majority of the sample with $\left\langle A_1^{v_z} \right\rangle_\mathrm{after}/\left\langle A_1^{v_z}  \right\rangle_\mathrm{before}<2$ meaning that their $m=1$ velocity field symmetry, whether weak or strong, was not affected by the passage of the perturber.  
The two bottom rows show two examples of the outlier 10 cases, where the change in $m=1$ amplitudes of the velocity fields were increased due to the flying-by Sgr-like perturbers. 
Fig.~\ref{orbits} shows the time evolution of the relative distance between hosts and satellites and the spatially averaged $\left\langle A_1^{v_z}  \right\rangle$ for the same five examples as in Fig.~\ref{examples}. The first example (ID=411449) has an already highly perturbed vertical velocity field, prior the considered pericenter. The orbit of the satellite has low eccentricity, so it is not surprising that no impact is seen after the pericentric passage, even if it filled the criteria defined in Section~\ref{sec:sample}. The other four examples have orbits with larger orbital periods, so the pericenters might have more impact than in the first case; however, only the last two examples (ID=414918 and 511920) present a noticeble increase in $\left\langle A_1^{v_z}  \right\rangle$. For ID=414918 the increase is also slightly delayed, which may be caused by the fact that the time of the pericenter obtained from the interpolated orbit is closely aligned with the snapshot, and the low cadence shows the increase in the velocity field only at the following snapshots that are saved already after $\sim100$ Myr. 

To find out what differentiates the cases where $m=1$ amplitudes were affected by the satellite passage from the others, we checked correlations with several parameters describing the properties of the interactions and the host galaxies. In Fig.~\ref{vz1} we plot the ratio $\left\langle A_1^{v_z} \right\rangle_\mathrm{after}/\left\langle A_1^{v_z}  \right\rangle_\mathrm{before}$ against the three parameters expected to be more important for interactions, i.e. the pericentric distance $D_\mathrm{peri}$, the ratio of the maximum circular velocities of the satellite to the host $v_\mathrm{c,sat}/v_\mathrm{c,max}$,
and the tidal parameter $S$. We estimated the pericentric distances using the 6D orbit interpolation following the prescriptions of \cite{Patton2024}. We slightly redefine the \cite{Elmegreen1991} tidal parameter $S$ by substituting the mass ratio with square of the $v_\mathrm{c,max}$ ratio, which yields 
\begin{equation}
    S=\bigg(\frac{v_\mathrm{c, sat}}{v_\mathrm{c, max}}\bigg)^2\bigg(\frac{R_\mathrm{host}}{D_\mathrm{peri}}\bigg)^3\bigg(\frac{\Delta T}{T}\bigg),
    \label{P}
\end{equation}
where $R_\mathrm{host}=5R_\mathrm{D}$ is the characteristic size of the host's disc, $\Delta T$ is the interaction time, defined as the time for the satellite to move one radian on its orbit around the pericenter, and $T$ is the time for the stars in the outer disc to move one radian in their orbits, approximated by \cite{Elmegreen1991} as $T=(R_\mathrm{host}^3/G M_\mathrm{host})^{1/2}$. We replace the mass ratio with $({v_\mathrm{c, sat}}/{v_\mathrm{c, max}})^2$, since the maximum circular velocity is a more robust proxy of the mass in cosmological simulations, as it does not depend on some scaling radius used to calculate the mass enclosed. Following the substitution of masses for circular velocities, in the approximation of $T$, we take $M_\mathrm{host}=(v_\mathrm{c,max}^2R_\mathrm{host})/G$.
The scatter plots in Fig.~\ref{vz1}, along with the overplotted Spearman correlation coefficients, show weak correlations, with $S$ exhibiting a slightly clearer trend.
The correlation with $S$ becomes stronger, although the sample size is smaller, when selecting the closest MW analogues. This may be due to two reasons. First, the fiducial sample of interactions was created by combining a relatively broad range of absolute parameter values for the MW analogues and a tighter range for Sgr-like interactions. Tidal $S$ is a dimensionless parameter, but it is possible that, in addition to dimensionless scaling, there exists an absolute scale relevant to this problem. The parameters of the closest MW analogues may better match this absolute scale, leading to amplified vertical velocity perturbations when combined with appropriately scaled interactions.
Second, the stronger correlation may simply indicate that $S$ does not encompass all the relevant physical ingredients, and other factors may also be contributing (such as resonances between frequencies of stars in MW disc and orbital frequency of the perturber, in analogy to results of \citealt{DOnghia2010,Lokas2010}) and be stronger in this subsample. The mild correlation between $S$ and resulting amplification of m=1 Fourier amplitudes of $v_z$ agrees with findings of \cite{Bennett2022} who notes the scaling between amplitude of vertical oscillations with the mass of Sgr.

Besides the interaction parameters, we looked whether the initial kinematical state of discs makes them more susceptible to Sgr-like perturbations. In Fig.~\ref{sigma_z} we plot the mean vertical velocity dispersions before (averaged in time and space the same way as $m=1$ Fourier amplitudes) against the mean vertical velocity dispersions after (top panel) and against after-to-before ratios of mean $m=1$ amplitudes (bottom panel). The mean velocity dispersions before and after correlate even stronger than the mean $m=1$ amplitude and also lie very close to the 1:1 line. This means that Sgr-like passages have little effect on vertical heating of the discs in our sample.\footnote{We checked whether different averaging in time and space of $\sigma_z$ changes this conclusion and found that it does not.} The bottom panel of Fig.~\ref{sigma_z} shows that $\left\langle A_1^{v_z} \right\rangle_\mathrm{after}/\left\langle A_1^{v_z}  \right\rangle_\mathrm{before}$ is greater than 2 mostly for galaxies that had their velocity dispersions smaller than 90-100 km/s, prior to the passage of the satellite. We conclude from this that Sgr-like passages can have a bigger impact on $m=1$ perturbations of the discs that were initially vertically colder. Moreover, in Fig.~\ref{sigma_v_fourier} we look whether cases that have $m=1$ warps induced by Sgr-like interactions are also getting heated and we find, with moderate correlation ($\rho=0.562$), that the outliers which experienced increase in the Fourier amplitude, also experienced heating in vertical velocities by the Sgr-like satellites.

The results presented in this subsection indicate that the vertical kinematics of MW-like discs in TNG50 on average are often already perturbed and interactions similar to the Sgr one have rarely an effect on them. Sgr-like interactions can have effect in few cases, where the disc is initially vertically cold. To test whether these results may be caused by the selection of the polar orbits, we modified the criterion 4 from Section 2 to create a control sample with orbits inclined with angles from the range of $0^\circ$–$60^\circ$. We found that only 6 out of 124 pericenter passages in this control sample had $\left\langle A_1^{v_z} \right\rangle_\mathrm{before}/\left\langle A_1^{v_z}  \right\rangle_\mathrm{after}>2$ and with the correlations showing very similar trends to the fiducial sample. This reaffirms our conclusions from the fiducial sample. Moreover, within the fiducial sample itself, we find that out of the 10 cases with the impactful effect of the satellite, 7 had orbits closer to prograde ($>90^\circ$) than retrograde ($<90^\circ$).

\section{Star formation}
\label{sect:SF}

\begin{figure}
\centering
    \includegraphics[width=.99\columnwidth]{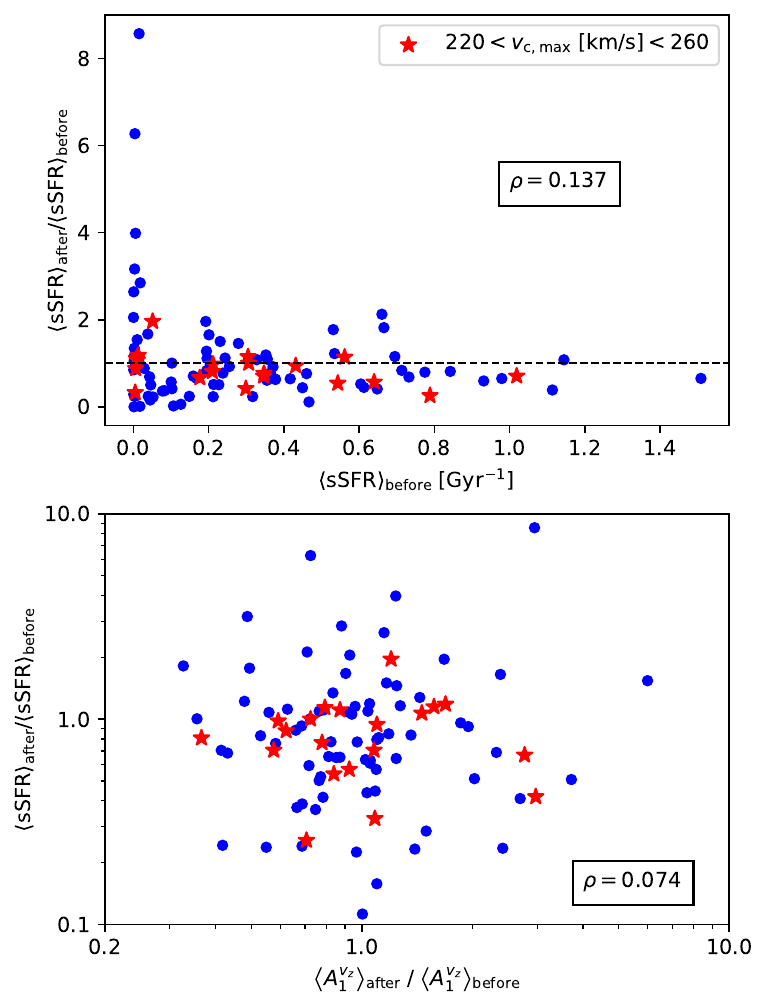}
    \caption{Top: Mean sSFR averaged over 6 snapshots before the pericenter passages vs after-to-before ratio of sSFR. Bottom: After-to-before ratio of sSFR against the after-to-before ratios of averaged $m=1$ Fourier amplitudes of $\overline{v}_z$. Red stars mark host galaxies with maximum circular velocities closest to the MW. The Spearman's correlation coefficient $\rho$ is indicated.}
    \label{sSFR}%
\end{figure}

\begin{figure}
\centering
\includegraphics[width=.85\columnwidth]{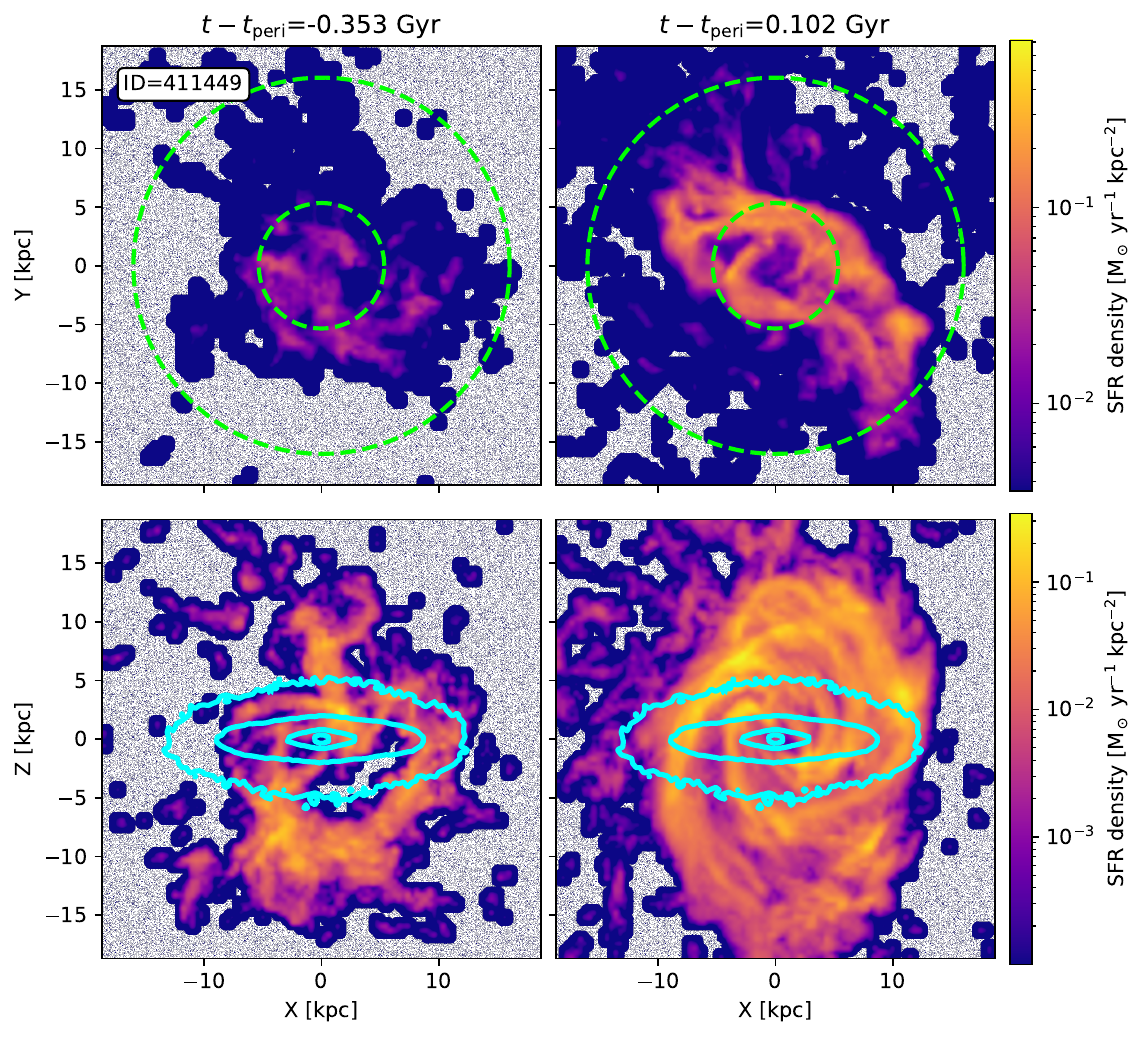}
\includegraphics[width=.85\columnwidth]{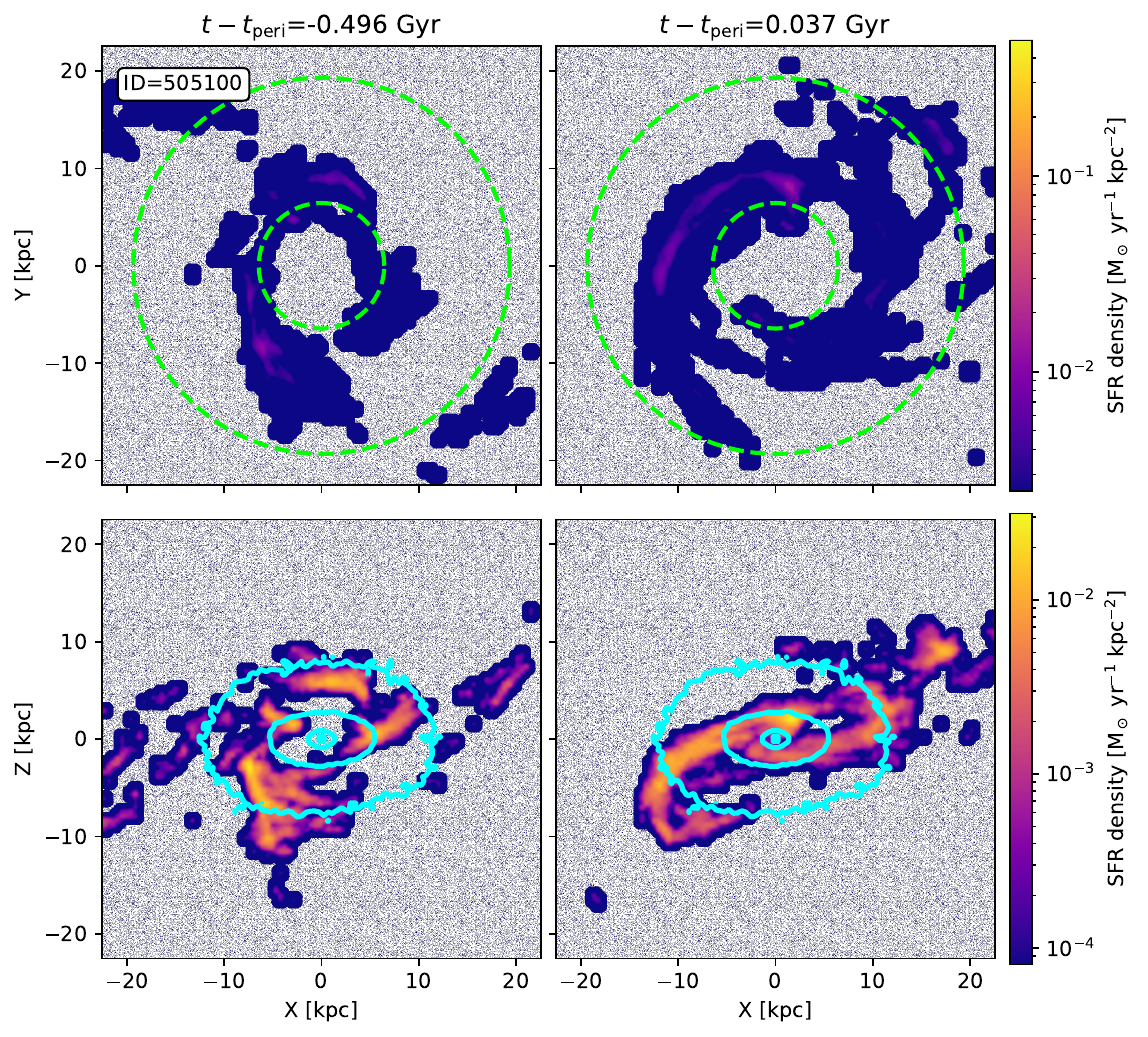}
\includegraphics[width=.85\columnwidth]{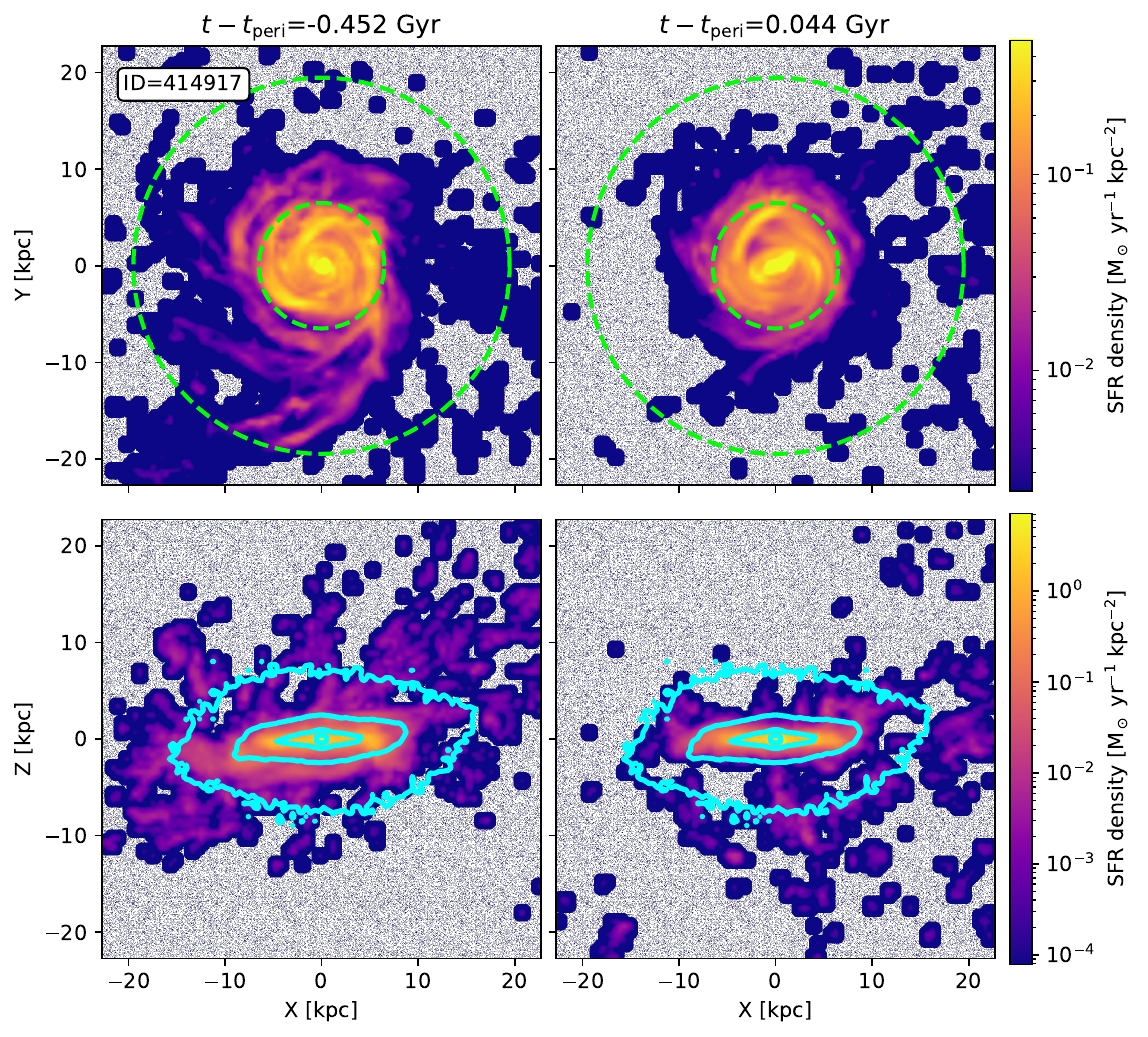}

\caption{Three examples of SFR density maps before and after the pericenter passages. In each group of four panels the top row shows face-on view and the bottom row shows edge-on. The time with respect to the pericenter passage $t_\mathrm{peri}$ is indicated in the title. Green dashed circles in the face-on maps mark the ring in which SFR was summed for the correlations in Section~\ref{sect:SF}. Cyan contours in the bottom rows mark density contours for the stellar disc, whose angular momentum was used to define the galactic plane.}
\label{SFR_0}
\end{figure}

Besides measuring the effect of Sgr-like perturbations on MW analogues on the vertical kinematics, we also looked into whether they can induce an increase in the star formation, as hypothesized for instance by \cite{RuizLara2020}. In order to quantify the possible increase in the SF, we do similar averaging as in Section~\ref{sect:kin}, that is at each snapshot we sum the star formation rates (SFR) of all gas cells in the radii between 2 to 6 $R_\mathrm{D}$ with the vertical cut of $|z|<2z_\mathrm{s}$. Later, we average these SFR in 6 snapshots before the pericenters and 3 snapshots after. Following results of \cite{Faria2025}, we normalize the SFR by stellar mass in these volumes to obtain the average specific SFR (sSFR) before $\left\langle \mathrm{sSFR} \right\rangle_\mathrm{before}$ and after $\left\langle \mathrm{sSFR} \right\rangle_\mathrm{after}$.  

The top panel of Fig.~\ref{sSFR} shows the after-to-before ratio of $\left\langle \mathrm{sSFR} \right\rangle_\mathrm{after}$ against $\left\langle \mathrm{sSFR} \right\rangle_\mathrm{before}$. The majority of cases in this plot scatter around the line where the ratio is equal to one, meaning that the passage of the satellite did not trigger any significant increase in the star formation in those cases. However, for cases with $\left\langle \mathrm{sSFR} \right\rangle_\mathrm{before}\simeq0.0\;\mathrm{Gyr}^{-1}$, several points lie on a straight vertical line with high values of the ratio (between 2-9). This indicates that the increase in the sSFR (which is mostly due to the increase of SFR, as we checked) happens when the host galaxy has little to no star formation before the satellite pericentric passage. 

We checked the SFR density maps for these cases with significant increases one by one to understand in a greater detail how this increase in the star formation works. The top two groups of four panels of Fig.~\ref{SFR_0} show two representative examples of these cases. The left columns show the face-on and edge-on SFR density distributions for 2 snapshots before the pericenters and the right ones for 1 snapshot after. The face-on maps confirm that there is indeed very little star formation happening in these galaxies before the satellite's passage. In these cases the edge-on maps show that this is often due to very perturbed/destroyed star forming discs (cyan contours outline the stellar disc density in the same reference frames). These quenched and highly perturbed star forming discs are likely caused by the strong kinetic AGN feedback, which is a known feature of the TNG50 simulation (e.g. \citealt{Semczuk2024}). 

On the other hand, the cases where there is little effect on the star formation from the satellite's passage have mostly well formed star forming discs, as in the example of the bottom panel in Fig.~\ref{SFR_0}. This example of the regular SF disc from the sample shows also that the SF discs are more compact than stellar discs (green lines come from exponential fits of stellar surface density). To check, whether these different scales affect the conclusion of the SFR amplification caused by Sgr-like passages, we looked into the correlations in the inner parts (0 to 2 $R_\mathrm{D}$) and found results to be consistent with the outer parts (2 to 6 $R_\mathrm{D}$).

The bottom panel of Fig.~\ref{sSFR} checks whether the increase in the star formation happens in the same cases that show the coincided increase in the $m=1$ Fourier amplitude in $v_z$. The lack of correlation shows that it does not, which is unsurprising, since for the star formation it happens mostly when the star forming disc is already very perturbed by the feedback, which does not directly affects the velocity field of the stellar disc.

\section{Discussion and summary}
\subsection{Vertical kinematics}
In this paper, we found that it is rare in the TNG50 simulations for Sgr-like satellite interactions to significantly impact the vertical kinematics of MW/M31-like discs. The initial sample of MW/M31 analogues \citep{Pillepich2024} and our defined sample of Sgr-like interactions span a broad range of masses, sizes, and initial kinematical states. 
In this multi-parameter problem, where host mass, perturber mass, pericenter distance, etc., all vary over wide ranges, it is likely that many combinations result in minimal impact. In contrast, the set of parameters that favor a strong impact is more limited, as it requires a specific alignment of conditions (e.g., high mass ratio, small pericenter, and a relatively cold disc prior to the interaction). One factor that may influence the resulting statistics is the declining stellar mass function of the TNG50 MW/M31 analogues (Fig. 4 of \citealt{Pillepich2024}), combined with the imposed upper limit of 50 kpc on pericenter distance. Since galaxy size generally scales with stellar mass, a fixed pericenter distance corresponds to a relatively deeper or more central encounter for more massive galaxies, while it remains comparatively more distant for lower-mass systems. Given that the majority of analogues are lower-mass galaxies, this mass-dependent effect could systematically reduce the impact of interactions near the 50 kpc threshold in the sample.

Interestingly, the host galaxies from our fiducial sample have relatively high values of $m=1$ Fourier amplitudes of vertical velocities, independently from the Sgr-like interactions (the median of $\left\langle A_1^{v_z} \right\rangle_\mathrm{before}$ is equal to 17 km/s). This implies that MW-like galaxies within this simulation are often vertically perturbed, which agrees with results from a smaller sample of galaxies from Auriga suite studied by \cite{Gomez2017}. It is difficult to determine, in all cases of our sample, the main driver of the $m=1$ signal in each of them (as was found for one case in the zoom-in study by \citealt{GC2024}); however, we can speculate on what predominantly perturbs them. In terms of tidal interactions, it could be that previous vertical disturbances persist for longer than the 1 Gyr time window used here to select isolated events. Alternatively, it could be the cumulative effect of many satellites below our mass threshold, which have therefore not been considered. \cite{Semczuk2020} found that around a third of the gas warps in TNG100 were induced by interactions, which, if extrapolated to stellar warps, suggests that other mechanisms may also be at play for the other cases. Warps can be triggered, for example, by misalignment between the disc and the halo \citep{Debattista1999}, or by misaligned \citep{Tigran2021,Tigran2022} or filamentary \citep{Arora2025} gas accretion. Another possibility is that the frequent warping of MW-like discs in this simulation is, to some extent, amplified by low stellar particle resolution. However, at $z=0$, the virial masses of MW/M31 analogues in \cite{Pillepich2024} all lie above the $10^{11}\mathrm{M}_\odot$ threshold for spurious heating estimated by \cite{Ludlow2023}. Moreover, \cite{SotilloRamos2023} argue that the flaring of MW/M31 discs spans a wide range of values compared to other high-resolution zoom-in simulations, further supporting the adequacy of the resolution for resolving the vertical structure of stellar discs. As always, in the case of resolution-related effects, further studies at higher resolution will be needed to confirm or refute the convergence of these results.

The observed frequency of warps, which is predicted here to be high in the redshift range 1 to 0, can be another test for cosmological simulations. With the growing data from instruments like JWST \citep{Reshetnikov2025}, it is now becoming possible to estimate warp fractions, similarly to the previously studied bar fraction.

\subsection{Star formation}
The results obtained here on the connection between MW-Sgr-like interactions and star-forming activity are strongly influenced by the AGN feedback prescription of TNG50. It was previously found that strong kinetic feedback can affect the mass-morphology relation \citep{Celiz2025}, the slowdown of bars \citep{Semczuk2024}, or their formation \citep{Lokas2022,Frosst2025}. In the case of the sample studied here, the feedback quenches star formation in the disc, and the analog of Sgr causes an increase only when the sSFR (and SFR) are very low (less than 0.1-0.2~$\mathrm{Gyr}^{-1}$). These values agree with the recent work of \citet{Faria2025}, who studied increases in sSFR in a sample of interacting galaxies in TNG100. Their sample, which experienced on average an amplification in sSFR of 1.6, also started with sSFR values around 0.14~$\mathrm{Gyr}^{-1}$.

The importance of feedback in addressing the research question posed here (can Sgr-like interactions trigger star formation bursts) highlights the difficulties of modelling star formation in simulations of galactic and cosmological scales. The formation of individual stars is not resolved, so the state of the star-forming gas results from various subgrid approximations. Whether the passage of a Sgr-like dwarf galaxy will affect the star-forming gas depends greatly on these prescriptions and how they shape the state of the gas disc. Other work on satellite induced SF bursts \citep{Bhargav2024, Li2025} highlight the complexity of this problem and dependence of the SF on many parameters.

Nevertheless, the fact that galaxies in our sample that experience sSFR amplification have been artificially quenched makes it interesting to relate to the fact that the MW is a relatively quiescent galaxy. In terms of numbers, \cite{Licquia2015} estimated the global sSFR to be $0.027\pm 0.006$~$\mathrm{Gyr}^{-1}$, implying that it is even lower in the outer disc, corresponding to the region where we conducted our measurements. This overlaps with the values of low sSFR where we found sSFR amplification to occur. While our Galaxy surely has not been quenched in the same way the AGN feedback destroys gas discs in TNG50, it exhibits low levels of sSFR that might make it prone to amplification caused by a passing satellite similar to Sgr. In other words, MW-like galaxies in TNG50 exhibit SFR amplifications coincided with a flying-by Sgr-like satellite, only when they are quenched. MW is a quiescent galaxy, however the quenching in TNG50 is not representative of the MW, since MW gas disc is still intact. 

\subsection{Summary}
In this paper we investigated, using TNG50 simulations, how interactions between MW-like galaxies and Sgr-like satellites affect the vertical kinematics and star formation of MW analogues. We searched the MW/M31 analogue sample in the redshift range 1--0 for isolated (within a time window of $\pm1$~Gyr) interactions with satellites of mass above $10^{10}\;\Msun$, pericenter passages under 50~kpc, and approximately polar orbits. Within this defined sample, only in 10\% of cases the vertical velocity fields of stellar discs were disturbed by the satellites. The majority of stellar discs already had a perturbed $v_z$ field (with a median amplitude of 17~km/s). The increase in $m=1$ perturbation of vertical kinematics coincided with the Sgr-like passage mostly in cold discs (with mean $\sigma_z<80-90$ km/s) and it mildly correlates with the strength of tidal interaction. In another 10\% of cases, the interaction coincided with increased star-forming activity. However, a significant amplification of sSFR happened only when the sSFR before the interaction was at low levels ($\lesssim0.15\;\mathrm{Gyr}^{-1}$), which in many cases was caused by prior intense AGN quenching, which is not representative for the MW. Sgr-like interactions can have an effect on MW-like discs in certain rare specific conditions. However, some of them, like quiescent SF discs, are created in TNG50 in a spurious way, which calls for future studies with different subgrid prescriptions.

\begin{acknowledgements}
We are grateful to the IllustrisTNG team for making their simulations publicly available. We acknowledge insightful discussions with J. Amarante, D. Patton and members of the GaiaUB team. This work was partially supported by the Spanish MICIN/AEI/10.13039/501100011033 and by "ERDF A way of making Europe" by the “European Union” and the European Union «Next Generation EU»/PRTR, through grants PID2021-125451NA-I00 and CNS2022-135232, and the Institute of Cosmos Sciences University of Barcelona (ICCUB, Unidad de Excelencia ’Mar\'{\i}a de Maeztu’) through grant CEX2019-000918-M. CL acknowledges funding from the European Research Council (ERC) under the European Union’s Horizon 2020 research and innovation programme (grant agreement No. 852839).
\end{acknowledgements}

\bibliographystyle{mnras}
\bibliography{sag}

\end{document}